\documentstyle[sprocl]{article}

\input{psfig}

\def\Journal#1#2#3#4{{#1} {\bf #2}, #3 (#4)}

\def\PLB{{\em Phys. Lett.}  B}

\def\PRD{{\em Phys. Rev.} D}
\def\PR{\em Phys. Rep.}

\def\JMPE{{\em Int. J. Mod. Phys.} E}

\def\be{\begin{equation}}
\def\ee{\end{equation}}
\def\bea{\begin{eqnarray}}
\def\eea{\end{eqnarray}}
\newcommand{\ba}{\begin{array}}
\newcommand{\ea}{\end{array}}
\newcommand{\p}{\partial}

\newcommand{\nn}{\nonumber}
\newcommand{\lb}{\label}
\newcommand{\re}[1]{(\ref{#1})}
\newcommand{\ds}{\displaystyle}

\newcommand{\vP}{{\vec P}}

\newcommand{\bdm}{\begin{displaymath}}
\newcommand{\edm}{\end{displaymath}}

\begin{document}

\title{THE NON-EQUILIBRIUM DISTRIBUTION FUNCTION \\OF PARTICLES AND 
ANTI-PARTICLES CREATED \\IN STRONG FIELDS}

\author{S.~A.~SMOLYANSKY, V. A. MIZERNY, D. V. VINNIK,\\
 and A.~V.~PROZORKEVICH}

\address{Physics Department, Saratov State University,
\\410071, Saratov, Russia\\E-mail:tar@sgu.ssu.runnet.ru}%

\author{V. D. TONEEV}

\address{Bogoliubov Laboratory of Theoretical Physics,\\
Joint Institute for Nuclear Research, 141980 Dubna, Russia \\
E-mail:toneev@thsun1.jinr.ru}


\maketitle\abstracts{ We investigate the quantum Vlasov equation
with a source term
describing  the spontaneous particle creation in strong
fields. The back-reaction problem is treated by solving this 
kinetic equation  together with  the Maxwell equation
which determines 
the induced time-dependent electric field in the system. 
The evolution 
of distribution functions for  bosons and fermions is
studied numerically. 
We found that  the system shows a regular dynamic
behavior if the back-reaction is neglected.  But if the 
back-reaction is included, it is not the case and  some stochastic
 features  are clearly 
revealed  in the non-equilibrium distribution function.}

\section{Introduction}

 In recent years much attention has been devoted to the 
back-reaction (BR) problem  in  high-energy particle 
physics~\cite{G87}$^-\,$\cite{K98}
and especially in early cosmology~\cite{Birrel}. 
A flux-tube model \cite{G87} based on the Schwinger
mechanism of the vacuum particle creation by a strong 
field 
is a commonly used model for the dynamical 
description of multiple particle phenomena.
During the passed time this process of the spontaneous pair 
creation has been intensively  investigated
for the case of a given external field, electromagnetic or 
gravitational one~\cite{Greiner,Grib}.  In majority of 
 the BR studies, 
some phenomenological source is introduced into a kinetic equation  
in a close analogy with the exact Schwinger  result for a
constant electric field (e.g.~\cite{G87}). Recently, a more 
consistent derivation of the source term has been done in 
Refs.~\cite{K98,Rau,s1}.
In particular,  an exact kinetic equation for scalar and
spinor QED with the non-Markovian source term for the  
time-dependent but 
spatially homogeneous field  was obtained in Ref.~\cite{s1}. Some 
properties of the kinetic equation of this type were studied 
in~\cite{s2}$^-\,$\cite{s4}.

A common feature of this approach is the observation of a very 
complicated pattern of oscillations in the  density of created 
particles. 
The frequency of these oscillations is of  order of  the 
 zitterbewegung  frequency~\cite{K93}
what evidences a high particle density reached in a 
system. 
It is quite obvious that under such conditions the interactions of 
created
particles should be taken into consideration. On the level of 
self-consistent
mean-field, it is just  the back influence of the created particles 
on a
formed field (the back reaction problem). For the thermalization 
process to be
very actual for ultra-relativistic heavy-ion collisions,  the
account of  direct particle-particle
interactions is important. To simplify the problem in question,  
 the relaxation time approximation~\cite{Baym} is used 
traditionally.

Longer than 10-years story of studies towards these directions put 
more new
questions than gave answers. This is related to high computational 
complexity 
of the equations under consideration which are integro-differential 
and highly
non-linear ones. In addition, these equations need a 
renormalization 
procedure  to remove the logarithmic divergences in the observable 
densities of energy and current. Besides, the 
 standard relaxation time approximation 
turned out to be too rough and 
should be improved  for the proper treating of 
collisions~\cite{s5}.

This contribution addresses  a single specific  problem, namely, to 
study 
the influence of the BR on temporal evolution of the creation 
process and
manifestation of this dynamics in observable fields and
especially in 
 particle distribution functions.   For simplicity,  we
do not take into account the non-Abelian structure of
color electric fields and concentrate on the back-reaction problem 
in
application only to strong  electromagnetic fields. However,  the 
characteristics considered and the
region  of the field parameters used are of interest  for
the flux tube model applications to hadronic processes.  We 
restrict
ourselves to a simplest situation of a time-dependent 
space-homogeneous 
field.

\section{Basic equations}

Our approach to the BR problem is based on the following
exact Vlasov-like KE for the distribution function $f(\vP,t)$
\cite{s1,s2} 
\be\label{2.1}
\frac{\partial f(\vP,t)
}{\partial t}+eE(t)\frac{\partial f(\vP,t) }{\partial
P_3}=C(\vP,t)\,\,,
\ee
where $E(t)$ - a strong homogeneous
electric field \footnote{)~We use the units $\hbar = c = 1$
and the metric is chosen to be $g^{\mu\nu} = {\rm
diag}(1,-1,-1,-1).$ }), 
$C(\vP,t)$ is the dynamical source term 
describing the vacuum
creation and annihilation processes within the Schwinger
mechanism 
\be\label{2.3}
C(\vP,t) = \frac{W(\vP ,t,t)}{2}
\int_{-\infty}^t dt' W(\vP ,t,t')
\big[1\pm 2f(\vP,t')\big]\cos\theta (\vP,t,t').
\ee
Here use the notation the transition amplitudes
\be\lb{2.4}
W(\vP ,t,t')=eE(t')\frac{P(t,t')}{\omega^2 (\vP,t,t')}
\left(\frac{\varepsilon}{P(t,t')}\right)^{g-1}\ ,
\ee
with the kinetic 3-vector momentum
$\vP(\vP_{\perp} ,P_3),$ 
  the degeneracy factor $g$ and  
\bea
P(t,t')=P_3+e[A(t)-A(t')],\\
\omega^2 (\vP,t,t')=\varepsilon_{\perp}^2+P^2(t,t'),\quad
\varepsilon_{\perp}=\sqrt{m^2+P_{\perp}^2},\\
\theta (\vP,t,t')=2\int_{t'}^t dt''\omega (\vP,t,t'').
\eea
The KE \re{2.1} is a direct consequence of the
corresponding one-particle
equation of motion in the presence of a quasi-classical
electric field $A_{\mu}=(0,0,0,A(t)),$ $E(t)=-\dot A(t)$.

For the subsequent calculations it is convenient to use
the following  local form of the KE \re{2.1} \cite{s4,MT79}
\bea\lb{3.2}
\frac{\p f}{\p t}+eE(t)\frac{\p f}
{\p P_3} = \frac12 { W}\,v\ , \nn \\
\frac{\p v}{\p t}+ eE(t)\frac{\p v} {\p P_3}=
\ds \, W\,[1\pm 2  f]-2\omega\, u \ ,\\
\frac{\p u }{\p t}+eE(t)\frac{\p u}{\p
P_3}= 2\omega\, v\ , \nn \eea
where was introduced two auxiliary real functions \cite{MT79}
$u(\vP,t)$ and $v(\vP,t)$
with the initial conditions $f(t_0)=v(t_0)=u(t_0)=0 $ and 
$ W=W(\vP ,t,t),\, \omega=\omega (\vP,t,t).$

 In the mean-field approximation, the distribution
function $f(\vP,t)$
allows one to find the densities of observable physical
quantities. In
particular, the conduction $j_{cond}(t)$ and polarization
$j_{pol}(t)$ terms  contribute into
the electromagnetic current density \cite{Grib}
\bea\lb{2.7}
j_{in}(t)=j_{cond}(t)+j_{pol}(t)\ ,\lb{2.6}\\
j_{cond}(t)=2eg\int\frac{d^3
P}{(2\pi)^3}\frac{P_3}{\omega}
f(\vP,t)\ , \\
j_{pol}(t)=eg\int\frac{d^3 P}{(2\pi)^3}\frac{P_3}{\omega}
\ v(\vP,t) \ 
\left(\frac{\varepsilon}{P_3}\right)^{g-1}\,.
\eea
The KE \re{2.1} should be combined with the Maxwell
equation
\be\lb{2.9} \dot E(t)=-j_{tot}(t)\ee
which closes  the set of equations for the BR problem. We
assume that a particle-antiparticle plasma was initially
formed due to 
some
external field $E_{ex}(t)$ excited by an external current
$j_{ex}(t)$. The internal field and current are noted as
$E_{in}(t)$
and $j_{in}(t)$. So, we have
\be\lb{2.10} E(t)=E_{in}(t)+E_{ex}(t)\,,\qquad
j_{tot}(t)=j_{in}(t)+j_{ex}(t)\,.
\ee

It is well known  that vacuum expectation values of type
\re{2.7}
can have  ultra-violet divergences and  need some
regularization
procedure. We use here the method suggested in paper
\cite{MT79}. To regularize  different observables
(currents, energy density etc.) it is
necessary to fulfill some subtractions of the relevant
counterterms from the every regularized function
$f,v$ and $u$. These subtraction terms are constructed
as  coefficients of the asymptotic expansion of corresponding
functions in series over the power of $\omega^{-1}(\vP)$.
The leading
terms of such expansions can be easily found from
Eqs. \re{3.2}
 \be\lb{3.4} f_a=
\left[\frac{eE(t) P_3}{4 \omega^3}
\left(\frac{\varepsilon_{\perp}}{P_3} \right)^{g-1} \right]^2,
\quad
v_a=e\dot E(t)\frac{P_3}{4 \omega ^4} \left(
\frac{\varepsilon_{\perp}}{P_3}\right)^{g-1}
 \lb{3.6}\ee
The conduction current \re{2.7} is regular one while the
polarization current \re{2.7}
contains the logarithmic divergence. For its
regularization it is
enough to fulfill one subtraction $v\to v-v_a$ in
\re{2.7}  that can be
interpreted as the charge renormalization. As the final
result,   the
regularized Maxwell equation can be written in the
following form
(it is implied here that the coupling constant $e$ and
the fields $E_{in}$
and $E_{ex}$ have been renormalized too) \cite{s4}:
\be\lb{3.7}
\dot E_{in}=- \frac{g e }{4\pi^3}\int d^3P
\frac{P_3}{\omega}\left[%
f+\frac{v}{2}
\left(\frac{\varepsilon_{\perp}}{P_3}\right)^{g-1}\!-e\dot E
\frac{P_3}{8\omega^4}\left(
\frac{\varepsilon}{P_3}\right)^{2(g-1)}%
\right].
\ee

This regularization procedure is  mostly adequate
to the method based
on  reducing the  KE to the system of partial
differential equations
\re{3.2}. This approach leads to the numerical results
(see \cite{s4})
which are quite consistent with the
 results obtained with  the adiabatic regularization
scheme
\cite{K93,K98}.

\section{Numerical results}

 The distribution function of a strongly
non-equilibrium state of particle-antiparticle plasma is
investigated numerically for the cases with ($E_{in}(t)~\ne~0$) and
without ($E_{in}(t)=0$) taking into account the BR mechanism.
 Various shapes  of the external field impulse were studied.
  It turned out that 
 the system behavior is weakly sensitive to a particular shape of 
the impulse and below  we present here the calculation 
results only for the impulse of the Narojny-type
\be\lb{impulse}
 A_{ex}(t)=A_0\, b\, [\tanh(t/b)+1], \,\qquad
E_{ex}(t)=A_0\,
\cosh^{-2}(t/b)\,.\ee
The parameters of this potential are chosen in accordance 
with conditions of the flux-tube
model \cite{K93}. In particular,
the coupling constant is taken as a rather large value,
$e^2=4$ and the used
dimensionless variables are~: $t\to t m , P\to P/m,
A\to A/m .$
The initial impulse is characterized by the  width
$bm=0.5$ and amplitude $A_0/m=7.0$.

\begin{figure}[h]  
\hspace{5mm}\fbox{ 
\psfig{figure=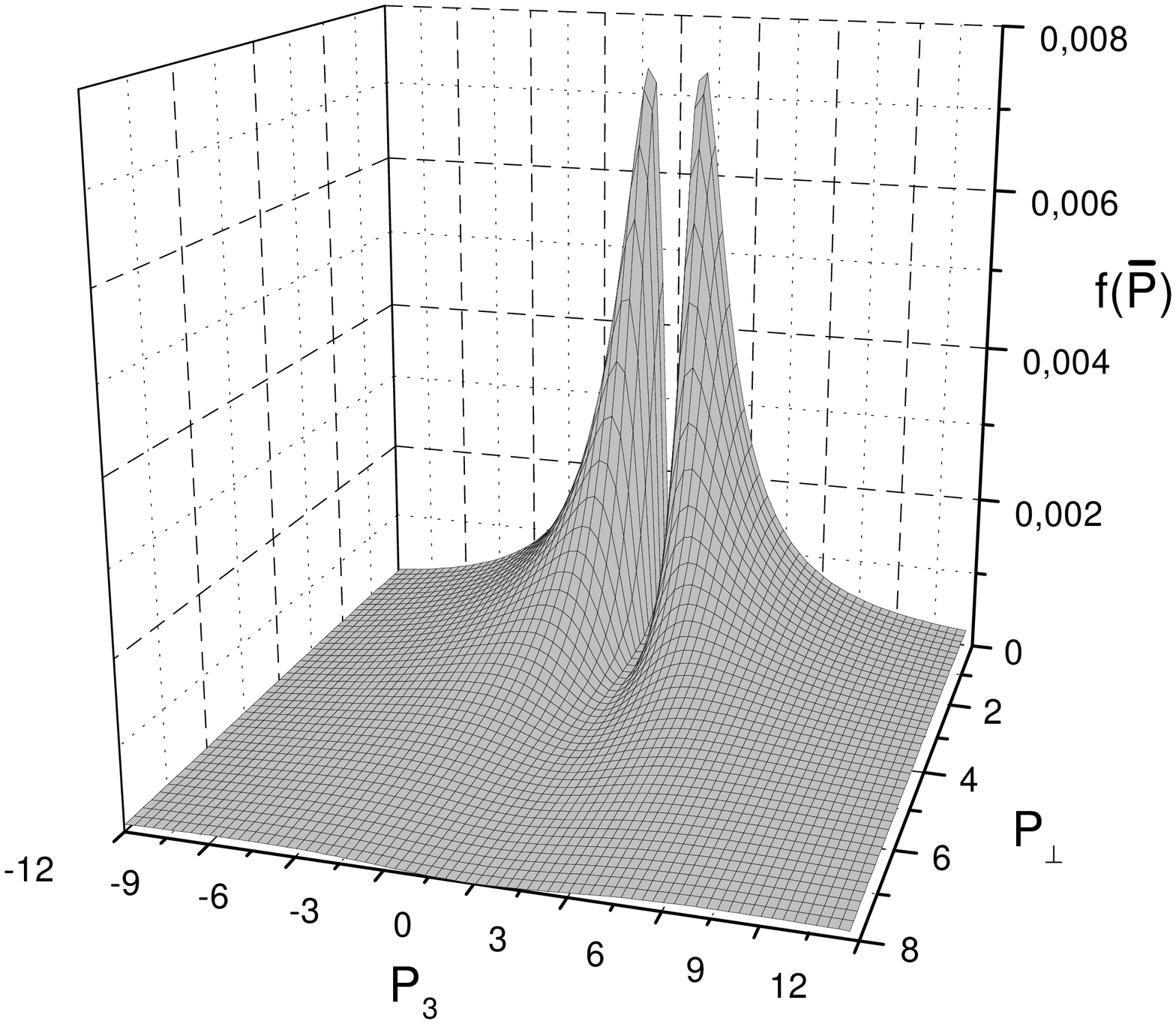,width=48mm,height=40mm,angle=0}}
\hspace{5mm} \fbox{
\psfig{figure=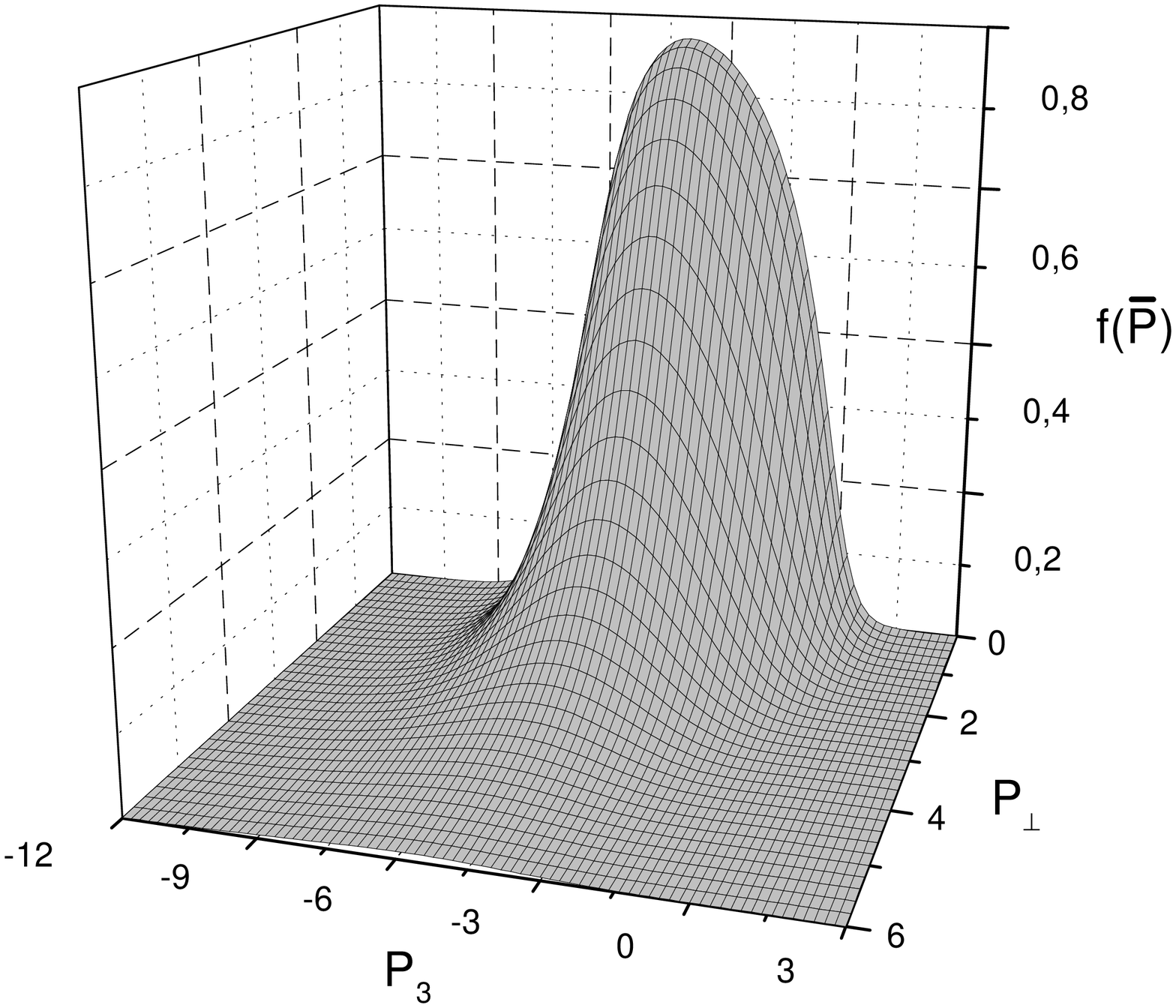,width=48mm,height=40mm,angle=0}}
\caption{Momentum distribution of produced bosons (left)and 
fermions 
(right) at time $t=0.05$. The influence of the back reaction is not 
included.}
\label{1}
\end{figure}

The distribution  functions of bosons and fermions  are shown 
in Fig.~\ref{1} neglecting  the BR mechanism.
 Because in this case their momentum dependence  is determined 
only by the transition amplitudes \re{2.4}, these 
distributions are smooth. Such behavior is regarded as a regular 
one.
A valley  region in the boson distribution function  arising 
near small values of $P_3$ is  caused by the
linear $P_3$-dependence of the amplitude \re{2.4}.

\begin{figure}[h]
\hspace{5mm}\fbox{
 \psfig{figure=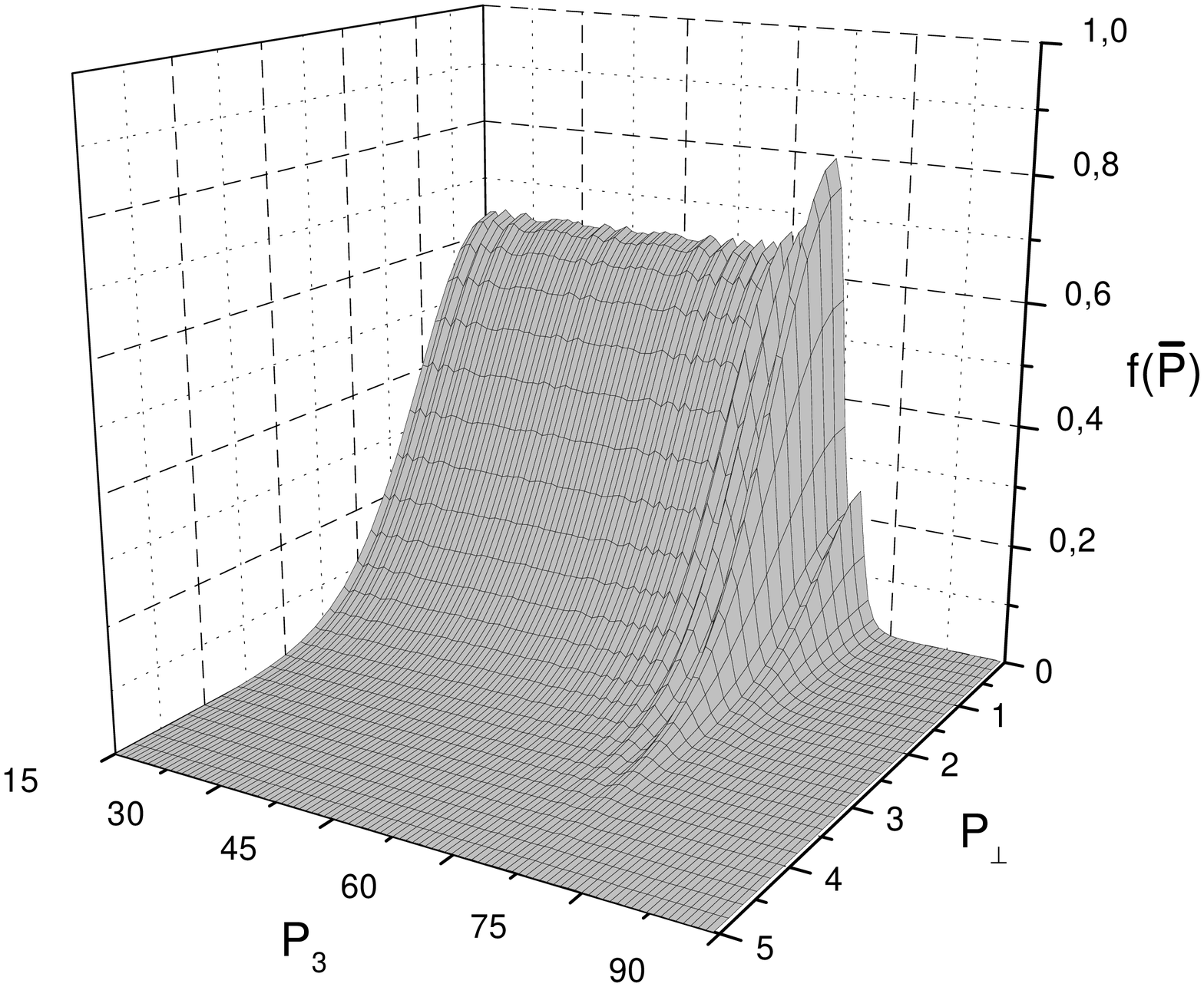,width=48mm,height=45mm,angle=0}}
\hspace{5mm} \fbox{
\psfig{figure=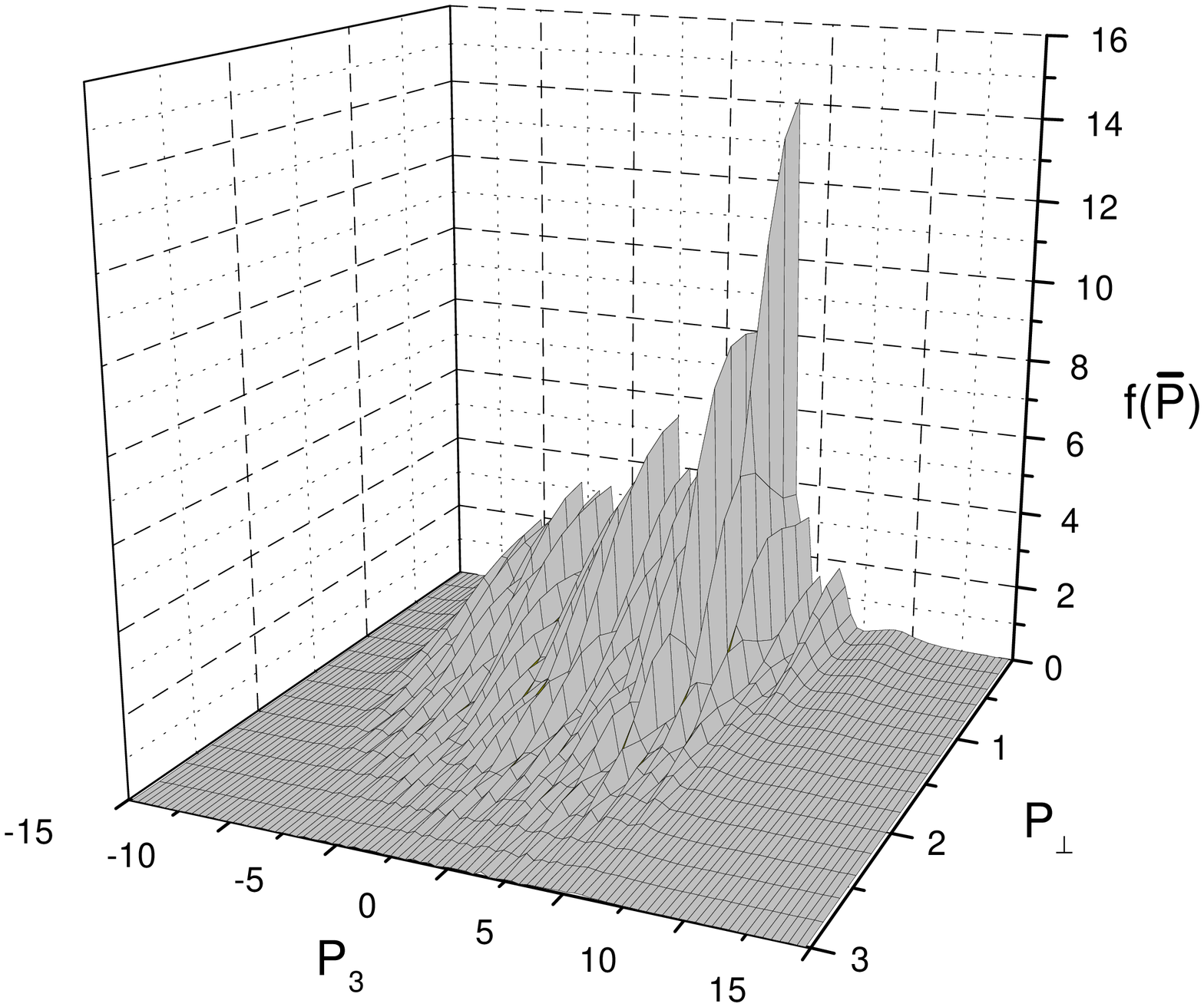,width=48mm,height=45mm,angle=0}}
\caption{Momentum distributions of produced bosons without (left) 
and
 with inclusion (right) of  the back reaction at time $t=10$.}
\label{2}
\end{figure}

\begin{figure}[h]
\hspace{5mm}\fbox{
 \psfig{figure=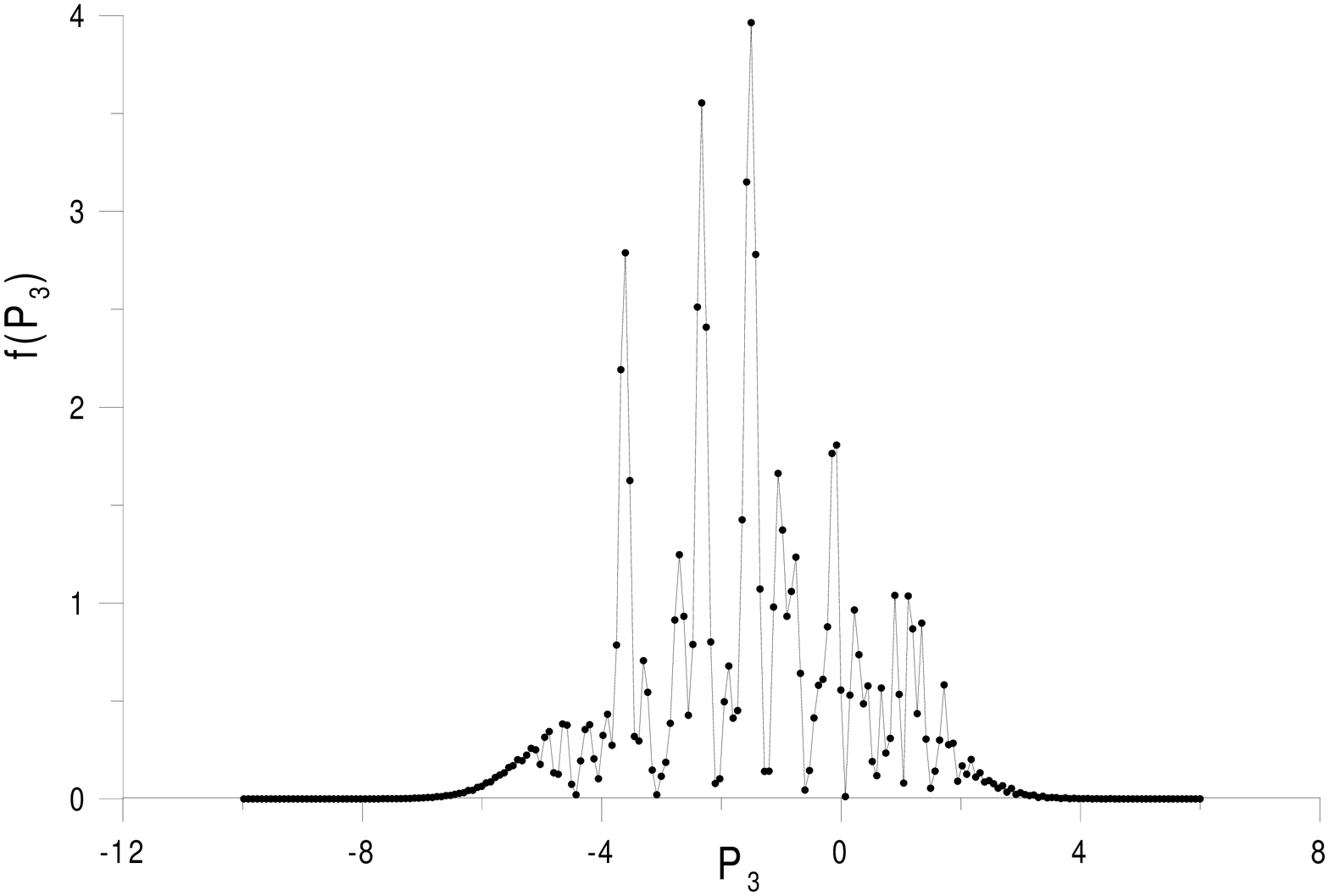,width=48mm,height=48mm,angle=0}}
\hspace{0.5cm} \fbox{
\psfig{figure=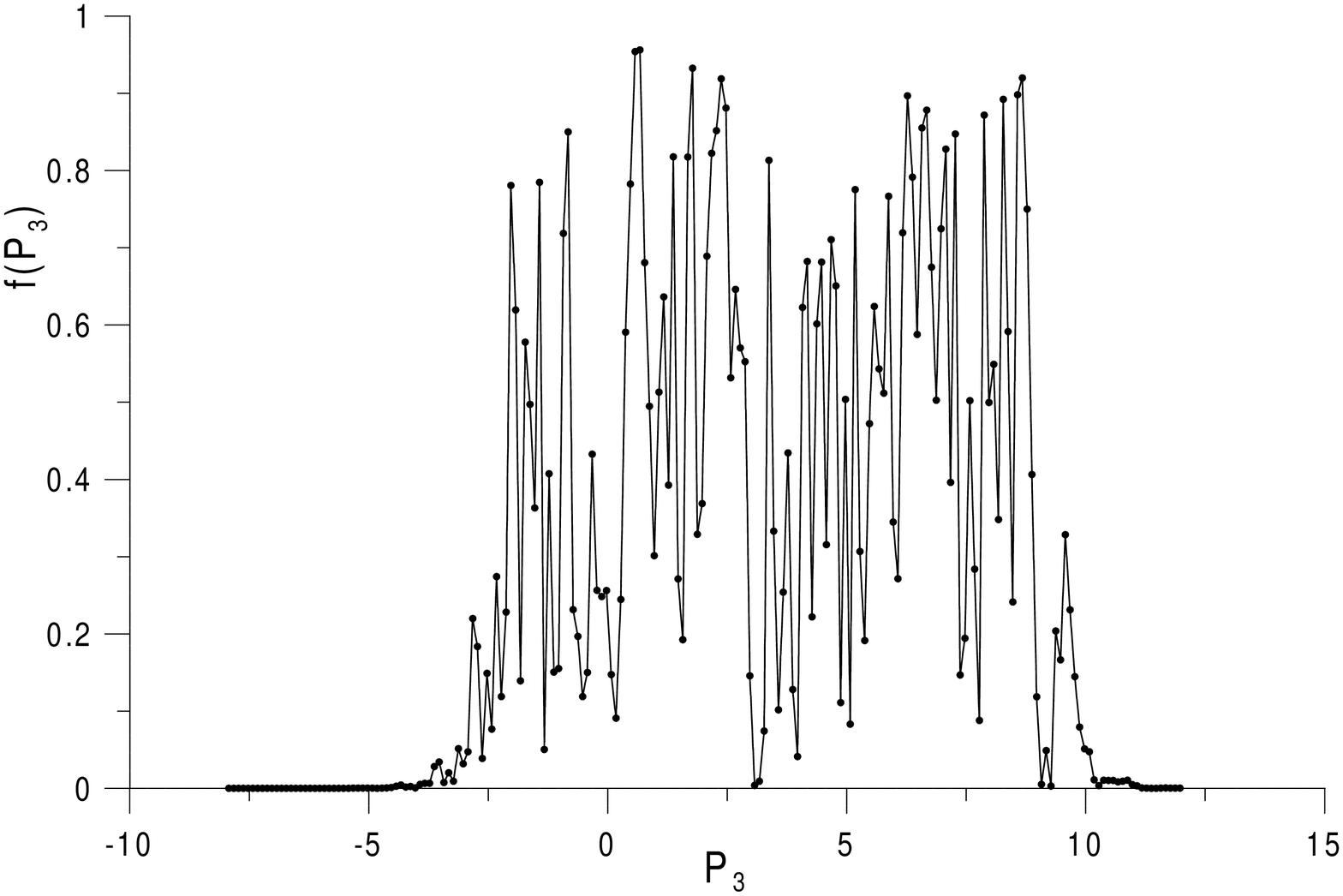,width=48mm,height=48mm,angle=0}}
\caption{Momentum distributions of produced bosons (left) and 
fermions (right)
at time $t=25$. Transversal momentum $P_\perp=0$ in both cases.}
\label{3}
\end{figure}

\begin{figure}[h]
\hspace{5mm}\fbox{
 \psfig{figure=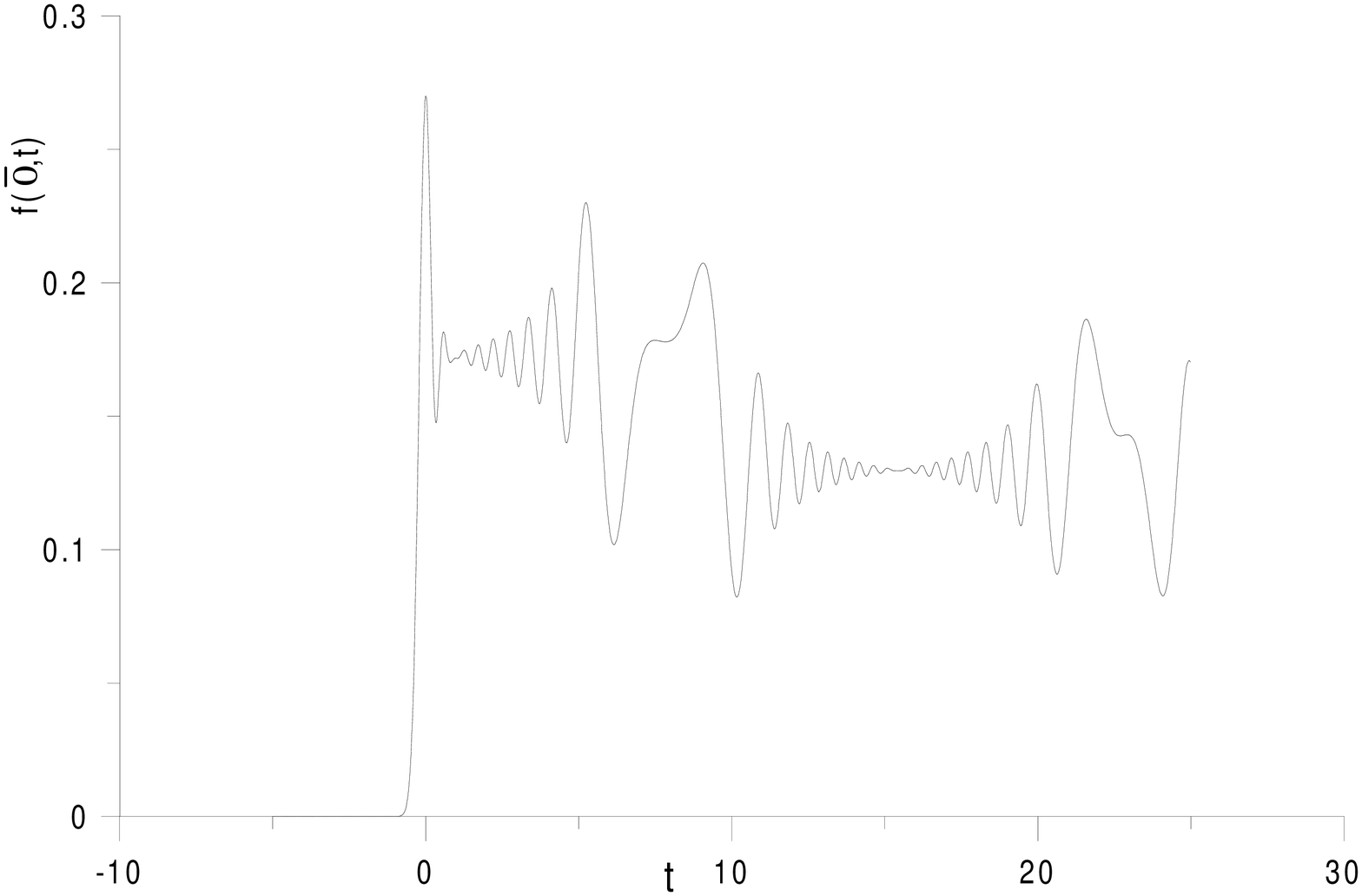,width=48mm,height=48mm,angle=0}}
\hspace{0.5cm} \fbox{
\psfig{figure=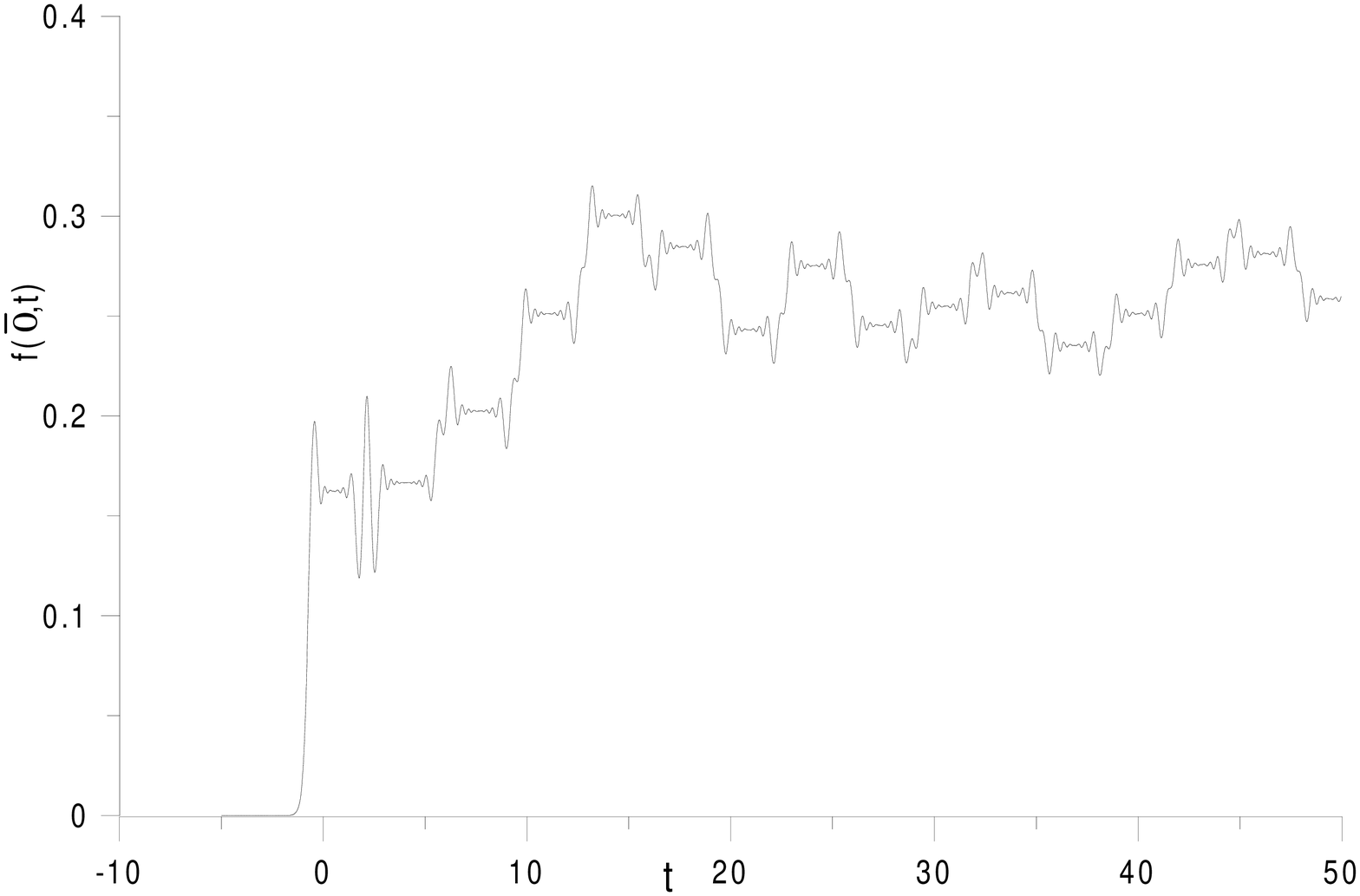,width=48mm,height=48mm,angle=0}}
\caption{Time evolution of boson (left) and fermion (right) 
distribution
function $f(\vec 0,t)$.}
 \label{4}
\end{figure}

\begin{figure}[h]
\hspace{5mm} \fbox{
 \psfig{figure=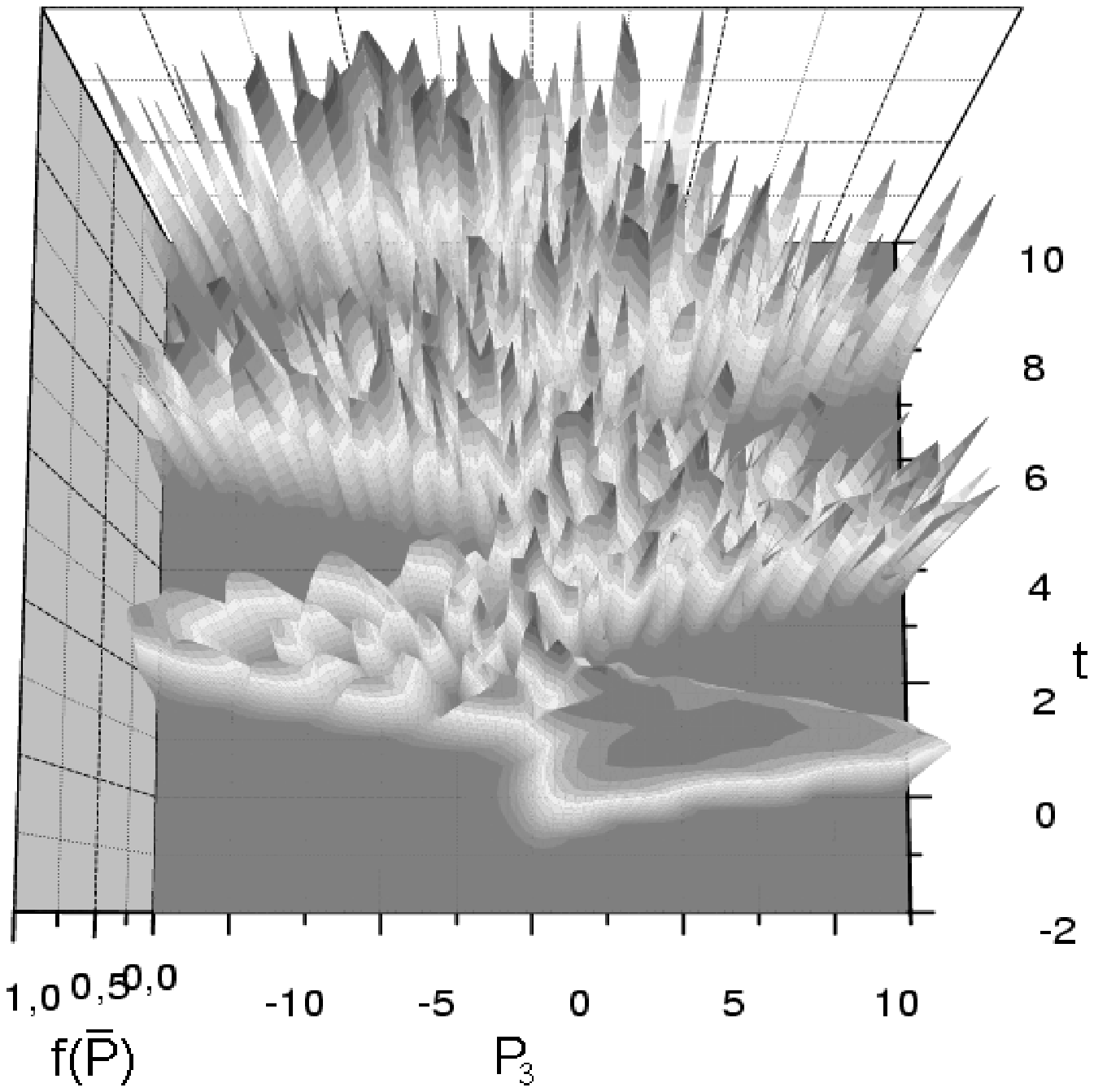,width=48mm,height=48mm,angle=0}}
\hspace{5mm} \fbox{
 \psfig{figure=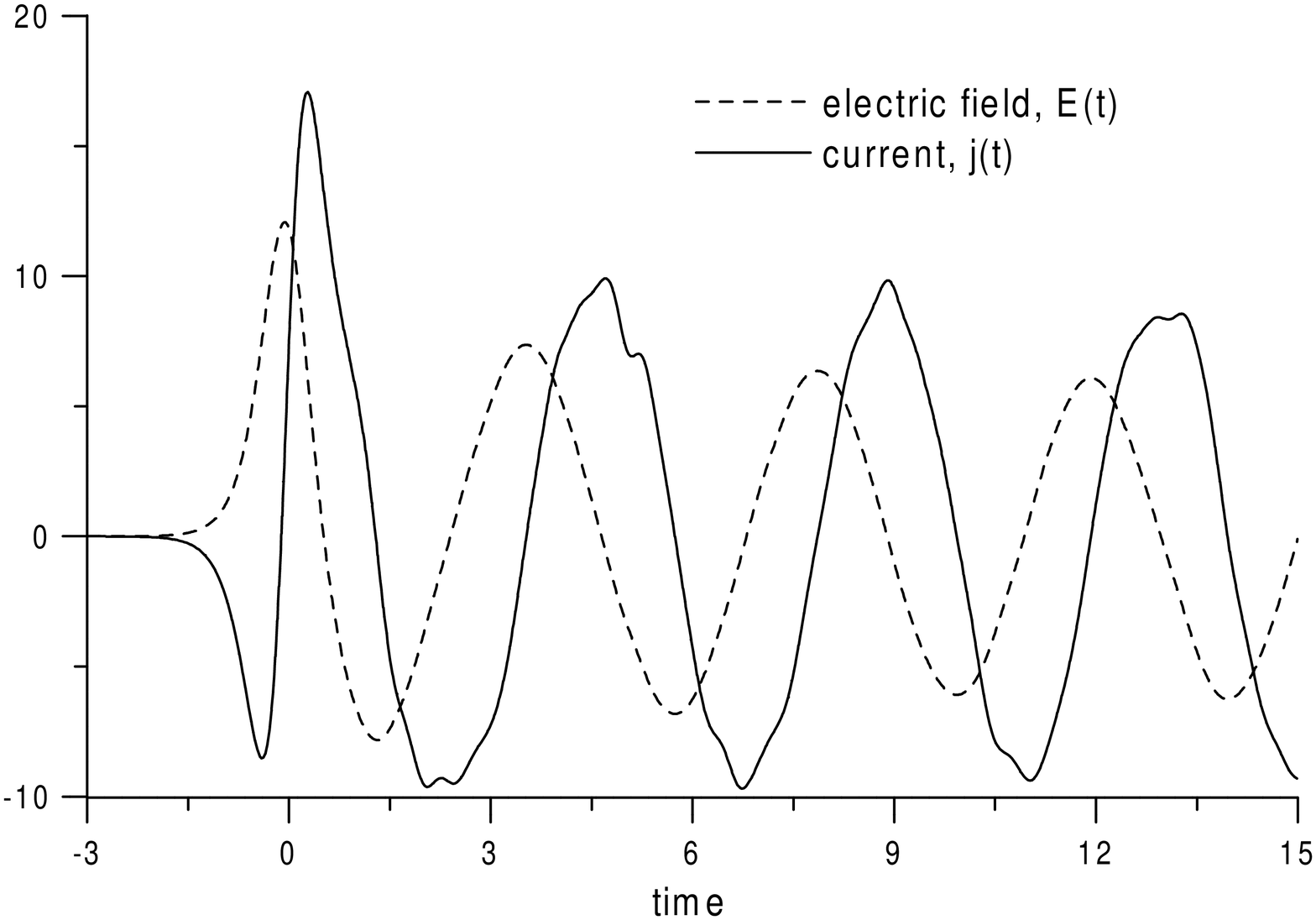,width=48mm,height=48mm,angle=0}}
\caption{Time evolution of $f(0,P_3,t)$  (left),  
electrical field and current (right) for the fermion case}
\label{5}
\end{figure}

 When
the  BR mechanism is taken into account, the regular momentum 
dependence of the distribution function is destroyed
(Figs.~\ref{2}-\ref{5};
the result like that in  Fig.~\ref{3} was obtained
previously in
\cite{K93} but in the framework of different
approach).

At the
same time Fig.~\ref{4} and Fig.~\ref{5} demonstrate the
existence
of periodic temporal behavior of the distribution
function.
Two-dimensional representation in Fig.~\ref{5} clearly
shows how
a 'dog-brush' structure of the distribution function
along the
$P_3$ axis is combined with the periodic time structure
along the
time axis.


\clearpage

\section{Conclusion}
Thus, the BR equations generate some large-scale
structure on the
background of small-scale multi-mode complex dynamics.
The small-scale 
trembling is a manifestation of vacuum oscillations.
Trembling 
frequency is increasing with time. The smoothed  initial
 distribution function in Fig.~\ref{5}
corresponds to the external field impulse. Large-scale
wiggles in the distribution function is a consequence of 
self-organization of the system \cite{Hak} due to the
growth of the
collective plasma oscillations.

The presented results provide some evidence that the
inclusion of
 the BR mechanism  into consideration of the pair
creation 
gives rise to stochastic 
behavior of the system. It is not excluded that a
dynamical chaos 
can be found also at other values of the parameter $eE$, 
however this  needs additional numerical investigations. 
It is quite possible that the revealed stochastic
features 
of the vacuum pair creation  is a source of the
statistical behavior
observed in multiple particle production.

\section*{ Acknowledgments}
One of the  authors (S. A. S.) gratefully acknowledges the 
hospitality
of the University of Rostock. He wishes to thank  G. R\"opke,
D. Blaschke, V. G. Morozov and S.M. Schmidt for valuable comments. 
This work was supported in part by the Russia State Committee of 
Higher 
Education under grant N~97-0-6.1-4.

\end{document}